\documentclass[12pt]{iopart}
\usepackage{epsfig}
\usepackage{amssymb}
\usepackage{color}
\usepackage{soul}
\usepackage{verbatim}
\soulregister\cite7
\soulregister\ref7
\newcommand{\tg}{{\rm g}}
\newcommand{\te}{{\rm e}}
\newcommand{\cD}{{\mathcal D}}
\newcommand{\cH}{{\mathcal H}}
\newcommand{\cL}{{\mathcal L}}
\newcommand{\ket}[1]{|{#1}\rangle}
\newcommand{\bra}[1]{\langle{#1}|}
\newcommand{\ketbra}[2]{|{#1}\rangle\langle{#2}|}
\newcommand{\braket}[2]{\langle{#1}|{#2}\rangle}
\begin{document}
\title{Protecting coherence by environmental decoherence: A solvable paradigmatic model}
\author{Juan Mauricio Torres$^{1,2}$ and Thomas H. Seligman$^{3,4}$}
\address{$^1$Instituto de F\'isica, Benem\'erita Universidad Aut\'onoma de Puebla, Apartado Postal J-48, Puebla 72570, M\'exico}
\address{$^2$Institut f\"{u}r Angewandte Physik, Technische Universit\"{a}t Darmstadt, D-64289, Germany}
\address{$^3$Instituto de Ciencias F\'isicas, Universidad Nacional Aut\'onoma de M\'exico, Cuernavaca, Morelos, M\'exico}
\address{$^4$Centro Internacional de Ciencias A. C., Cuernavaca, Morelos, M\'exico}

%\pacs{03.65.Yz}{Decoherence; open systems}
%\pacs{42.50.Pq}{Cavity quantum electrodynamics}

\date{\today}

\begin{abstract}
We consider a particularly simple exactly solvable model for a qubit 
coupled to sequentially nested environments. The purpose is to exemplify 
the coherence conserving effect of a central system, that has been reported as a result 
of increasing the coupling between near and far environment. 
The paradigmatic example is the Jaynes-Cummings Hamiltonian, 
which we introduce into a Kossakowski-Lindblad master equation using 
alternatively the lowering operator of the oscillator or its
number operator as Lindblad operators. 
The harmonic oscillator is regarded as the near environment of the qubit, while
effects of a far environment are accounted for by the two options for the 
dissipative part of the master equation.
The exact solution allows us to cover the entire range of coupling strength from 
the perturbative regime to strong coupling analytically.
The coherence conserving effect of the coupling to the far environment is confirmed 
throughout.

\end{abstract}
\maketitle

\section{Introduction}

Decoherence was and is not only a central theme of physics, but it is also a central problem for any practical implementation of  quantum 
computation and quantum information schemes \cite{Nielsen}.
%The loss of coherence in quantum systems is one of the major obstacles to overcome 
%in order to implement quantum technologies such as quantum computation and quantum information protocols \cite{Nielsen}. 
The source of decoherence is the surrounding environment to which the system under investigation couples invariably.
Dynamical decoupling is a well established technique to isolate a physical system or to tailor 
a desired Hamiltonian evolution \cite{Viola98,Viola99}. Other dynamical control methods exploit the quantum Zeno effect  to slow down decoherence processes \cite{Kurizki,Kofman,Wu}.
In these techniques, one of the requirements is a periodic driving or measurement of the system.
A natural question is whether intrinsic decay mechanisms of the environment can enhance the coherence of a central system.
Recent studies have considered the coherence loss of a ``near'' 
environment to stabilize the coherence of a central quantum system.
This reaches from the limit of very fast decoherence of the near environment \cite{Zanardi,MGS,GMS,Gonzalez,Man}, 
which may actually lead in some limit to a protected subspace for the central system,
all the way to perturbative treatments where all couplings are small
\cite{MGS, GMS}. 
The great benefit in this approach is that one does not require to control the system dynamically.
Numerical results for spin systems \cite{Gonzalez} and random matrix 
environments \cite{MGS} seem to support such improvement throughout the range of coupling strength. 
In all cases there seem to exist options to improve the persistence of coherence  of the central system 
considerably if it is already quite good to start with. 
Indeed it so seems, that weak coupling of the central system to the near environment 
is the only prerequisite for this method to be workable. The results are positive and interesting but a little counter intuitive.

Under such conditions an exactly solvable example is usually very enlightening and that is what we are going to present. 
We will use a fairly new technique to obtain exact solutions of the corresponding Kossakowski-Lindblad master equation \cite{Torres2014}. 
We shall thus deepen the understanding of the role of nested environments  for decoherence and simultaneously provide a non-trivial 
application of a new technique to solve open quantum problems. Our study focusses  on two versions of a paradigmatic, simple, 
and exactly solvable  model in quantum optics: A Jaynes-Cummings model with dephasing and 
a damped Jaynes-Cummings model. The dephasing and the damping mechanisms are
assumed to act solely on the cavity mode. The role of the central system is played by a two-level atom which 
is coupled to a single mode of an optical cavity acting as the near environment.
Decoherence of the cavity is taken into account to mimic the effects
of a far environment. We assume that the two types of decoherence mechanisms in the cavity which
can be described in terms of a Markovian master equation in 
Kossakowski-Lindblad form \cite{Gorini,Lindblad}.
In a first approach we consider a dephasing model, with the number operator as Lindblad
operator. This may not be very realistic, but it will turn out to be most illustrative
due to its simple analytical treatment and the absence of competing effects: The  Liouville operator can be expressed in terms of disconnected
$4\times 4$ matrices. 
In a second case, photon losses are considered by choosing the annihilation operator
as Lindblad operator. This case has deep roots in the field 
and can be connected to the standard setting in the 
Haroche experiment \cite{Haroche} and according to Garraway's pseudo mode theory \cite{Garraway,Dalton,Mazzola}, 
it is equivalent to a two level atom interacting with a continuum of modes.

%For both cases, we have worked out analytical expressions of the reduced density matrix
%of the atom and its purity to measure the decoherence of the central system.
The simple fact that we do get exact solutions in special but non trivial situations
for the decay of coherence 
of a rather complicated system is of great interest, as it will allow us to gain insight 
in possible mechanisms leading to protection of coherence in nested environments, 
that previous to this work were numerically detected and analytically derived for extreme situations.
%We find that in the strong coupling
%limit of the cavity to the outside world, the atomic state 
%remains pure, i.e., it evolves coherently in time. 

We shall start by briefly defining the model and then proceed to discuss the
simpler case, where the Lindblad operator is the number operator of the
harmonic mode. 
Next we will address the case of photon losses in the cavity in which the annihilation
operator of the oscillator is considered as Lindblad operator.
Finally we shall discuss to what extent our considerations may shed light 
into known numerical and random matrix results \cite{MGS,GMS,Gonzalez},
and discuss in what ways the basic result can be used to help control coherence.

\section{The model}
\begin{figure}[t]
\begin{center}
\includegraphics[width=.45\textwidth]{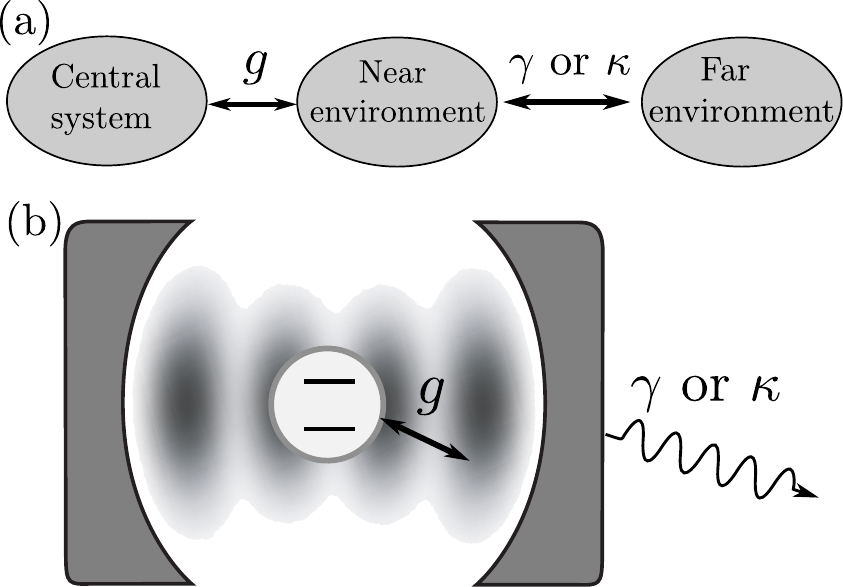}
\caption{\label{sketch} 
(a) Sketch of the nested environment setting: a central system couples
to a near environment with strength $g$. The near environment interacts with 
the far environment with an effective coupling parameter either $\gamma$ or $\kappa$.
(b) Sketch of the Jaynes-Cummings model: a two level atom (central system)
interacts with an optical cavity (near environment) with interaction strength $g$.
The cavity presents either dephasing at rate $\gamma$ or photon losses at
rate $\kappa$ (to a far environment). 
}
\end{center}
\end{figure}

The Jaynes-Cummings (JC) model describes the interaction between a two-level atom with one
mode of the electromagnetic field inside an optical cavity \cite{Jaynes1963}.
The Hamiltonian in  the interaction picture with respect to the 
electromagnetic field energy is given by ($\hbar=1$)
\begin{equation}
  H=
  \delta\sigma_+\sigma_-+  g(a\sigma_++a^\dagger \sigma_-),
  \label{Hamiltonian}
\end{equation}
where $\sigma_\pm$ are the raising and lowering operators 
of  the two-level atom acting on the Hilbert space $\cH_{\rm at}=\mathbb{C}^2$,
% (with the standard inner product),
while $a$ and $a^\dagger$ are the cavity mode creation and annihilation
operators that act on the Fock space $\cH_{\rm cav}=F_+(\mathbb{C})=\ell^2$ \cite{Fock}.
The complete Hilbert space is therefore the tensor product of the composite Hilbert spaces 
$\cH=\cH_{\rm at}\otimes \cH_{\rm cav}$. The parameter $g$ is the interaction strength 
between the two-level atom and the cavity, while $\delta$ is the detuning of the atomic transition frequency 
from the frequency of the mode.
A general state of the system can be represented by 
the density matrix $\rho$ that is an operator acting on $\cH$. Its Hamiltonian dynamics 
is governed by the von Neumann equation $\dot\rho=-i[H,\rho]$. We introduce decoherence effects
in the system
by adding the action on $\rho$ of the generator $\cD[A]$ defined in Lindblad form as
\begin{equation}
  \cD[A]\rho=A\rho A^\dagger-\frac{1}{2}\left(A^\dagger A \rho+\rho A^\dagger A\right).
  \label{dissipator}
\end{equation}
The Lindblad operator $A$  could in principle be chosen to act on the
composite Hilbert space $\cH$, as in the case of more realistic models that
consider combined decay mechanisms in Lindblad form \cite{Govia,Beaudoin}.
However, in this work we restrict ourselves to  operators
$A$ acting solely on the Hilbert space of the cavity $\cH_{\rm cav}$. 
The aim is to treat the two-level atom as a central system, the cavity as a near 
environment and the effects of a far environment described by the Lindblad operators. 
The model an its connection to the nested environment description is shown in Fig. \ref{sketch}. In the next two sections we will consider first a dephasing and then a photon loss operator.

\section{Dephasing of the cavity}
Let us start our discussion by considering a situation that involves a dephasing mechanism
in the cavity but without any loss 
of excitations.
The dynamics of a model that includes this effect can be 
described by the following  master equation 
\begin{equation}
  \dot\rho=\cL_{\rm d}\rho
  =-i\left[ H,\rho\right]+\gamma\cD[a^\dagger a]\rho
  \label{masterD}
\end{equation}
depending on the dissipator of Eq. \eref{dissipator} and with 
the Lindblad operator $A=a^\dagger a$. 
An important property of the Liouville operator $\cL_{\rm d}$ is that it 
conserves the number
of excitations of the operator $a^\dagger a +\sigma_+\sigma_-$. Therefore,
if one considers initial states of the form 
$\ket{\te,n-1}=\ket{\te}\otimes\ket{n-1}$, that is 
an excited atom and $n-1$ photons in the cavity, the time dependent 
density matrix of the system can be expressed as
\begin{eqnarray}
\fl\rho(t)=
v_1\ketbra{\tg,n}{\tg,n}+v_2\ketbra{\tg,n}{\te,n-1}
+v_3\ketbra{\te,n-1}{\tg,n}+v_4\ketbra{\te,n-1}{\te,n-1}
.
\end{eqnarray}
The time dependent coefficients 
are solution of the differential equation
$\dot{\vec v}=L\vec v$, where 
$\vec v=(v_1,v_2,v_3,v_4)^\top$ is a column vector and
$L$ is a matrix that can be obtained from the Liouvillian $\cL_{\rm d}$ and has the explicit form
\begin{equation}
  L=\left(
  \begin{array}{cccc}
    0&ig\sqrt n &-ig \sqrt n &0 \\
    ig\sqrt n&-\gamma+i\delta& 0 &-ig \sqrt n \\
    -ig\sqrt n &0&-\gamma -i\delta&ig \\
    0&-ig\sqrt n &ig \sqrt n &0 
  \end{array}
  \right).
  \label{Lmatrix}
\end{equation}
Actually, the matrix $L$ is one of many disconnected blocks that form the Liouvillan $\cL_{\rm d}$ \cite{Torres2014}.
As it is a $4\times 4$ matrix, the eigenvalues of $L$ can always be calculated in closed form as
shown in  \ref{appendixL}.
However,  we focus our attention to the resonant case as it captures the qualitative essence of the dynamics 
we want to describe and the resulting equations can be written in compact form. Deviations from this condition
do not present a qualitative change in the long time behaviour that we are interested.
Therefore, in the resonant case, i.e., $\delta=0$,  the four eigenvalues of $L$ are  given by
\begin{equation}
  l_0=0,\quad l_1= -\gamma,\quad l_\pm^{(n)}=
  -\gamma\frac{1\mp \sqrt{1-\eta_n}}{2},
  \label{evalLD}
\end{equation}
where we have introduced the dimensionless parameter
\begin{equation}
  \eta_n=16g^2n/\gamma^2.
  \label{eta}
\end{equation}
This parameter $\eta_n$ can be seen as a rescaled interaction strength between atom and cavity, that tends to zero
for increasing values of $\gamma$ whenever the values of $g$ and the photon number $n$ are finite.
\begin{figure}[ht!]
\begin{center}
\includegraphics[width=.45\textwidth]{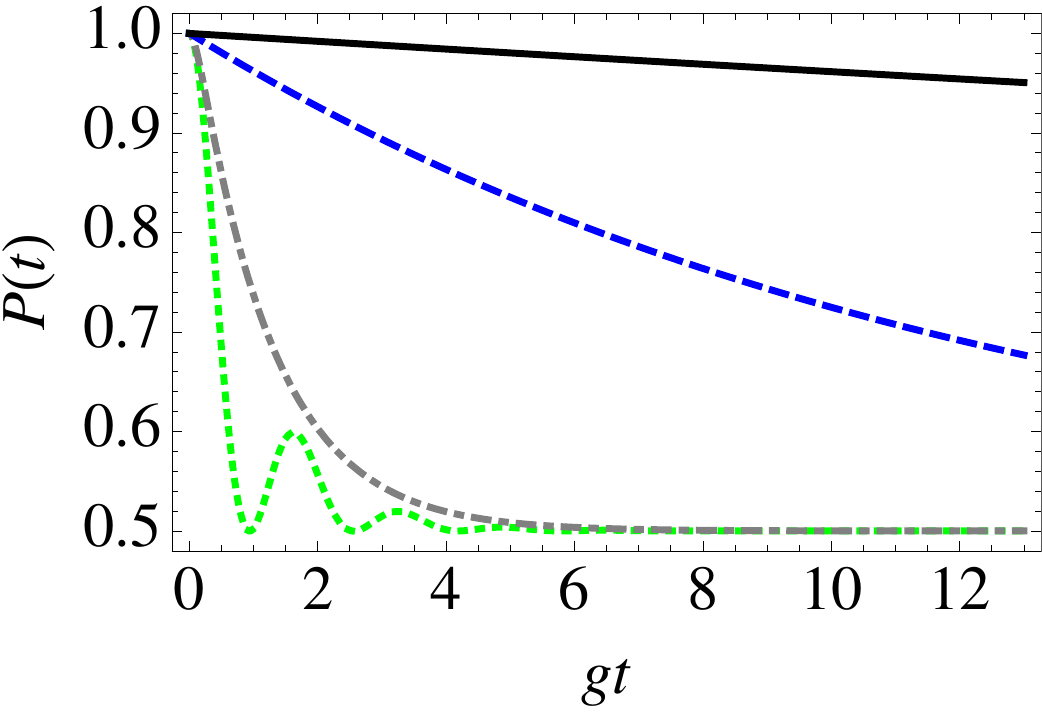}
\includegraphics[width=.45\textwidth]{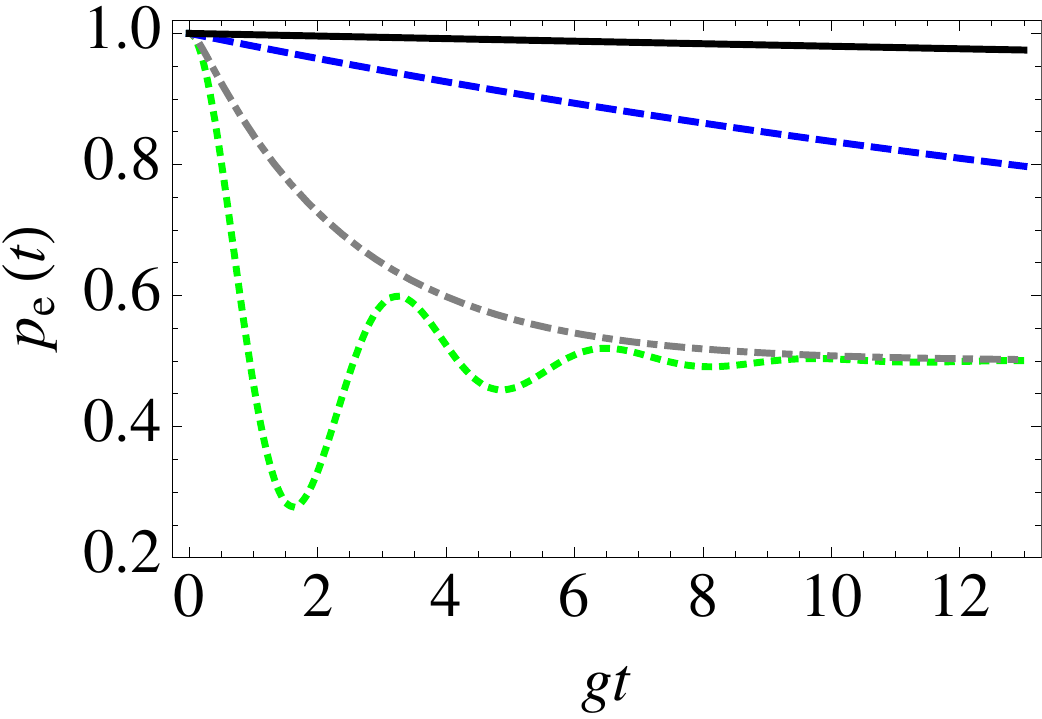}
\caption{\label{plotD} Left (right) plot: atomic purity $P(t)$ (excited state population  $p_{\rm e}(t)$) 
as a function of time in units of $1/g$ for different value of the 
cavity's dephasing parameter $\gamma$. 
The initial state is an excited atom and
an empty cavity subject to the dynamics of Eq. \eref{masterD}.
Dotted (green) line $\gamma/g=1$, dotted-dashed (grey) line $\gamma/g=10$,
dashed (blue) line $\gamma/g=100$, and full (black) line $\gamma/g=1000$. 
Detuning is $\delta/g=0.$ 
Note: The purity $P(t)$ is plotted in the range of minimum to maximum purity ($[0.5,1]$).
}
\end{center}
\end{figure}
As we are particularly interested in the coupled dynamics in this limiting case  
%values of the dephasing parameter 
($\gamma \gg g,n$),
we analyze the eigenvalues by expanding them in terms of $\eta_n$ which 
leads to the expression
\begin{equation}
  l_\pm^{(n)}=-\frac{1\mp 1}{2}\gamma-\frac{1}{4}\gamma\eta_n+\mathcal{O}[\eta_n^2].
  \label{expevalD}
\end{equation}
It follows from this expansion %in Eq. \eqref{expevalD} 
that the eigenvalues of $L$, which are also eigenvalues
of the Liouville operator $\cL_{\rm d}$, are insensitive to the coupling strength for sufficiently large values of
the dephasing parameter $\gamma$. 
This is already an indication that the atomic system is ``protected'' from the presence of the 
cavity by the dephasing mechanism. 

Now we turn our attention to the dynamical properties of the atomic sub-system.
By solving the differential equation and tracing over the photonic degree of 
freedom, one can evaluate the atomic density matrix which is diagonal in this case and is 
given by
\begin{equation}
  \rho_{\rm at}(t)=
  \frac{1}{2}\left[1-h_n(t)\right]\ketbra{\tg}{\tg}+
  \frac{1}{2}\left[1+h_n(t)\right]\ketbra{\te}{\te},
\end{equation}
were we have considered the complex function
\begin{equation}
  h_n(t)=%\sum_{\ell=\pm}
  \frac{\eta_n e^{l^{(n)}_+t}}{2\eta_n+4 l^{(n)}_+/\gamma}
  +\frac{\eta_n e^{l^{(n)}_-t}}{2\eta_n+4 l^{(n)}_-/\gamma}.
\end{equation}
In the limit of $\eta_n\to 0$ ($\gamma\gg g,n$), the first term tends to $1$ while the second tends to zero. 
This can be noted by taking into account the form of $l_\pm^{(n)}$ in Eq. \eref{expevalD}
and of $\eta_n$ in Eq. \eref{eta}.
Therefore, in this limiting case  the probability of finding the atom in the excited state 
\begin{equation}
  p_{\rm e}(t)=\Tr\left\{\sigma_+\sigma_-\rho_{\rm at}(t)\right\}=\frac{1+h_n(t)}{2}
\end{equation}
freezes at unit value. 
One can also consider purity of the atomic state. For this quantity one 
finds 
\begin{eqnarray}
  P(t)&={\rm Tr}\left\{\rho_{\rm at}^2(t)\right\}=\frac{1}{2}+\frac{\eta_n e^{-\gamma t}+f_n^+(t)+f_n^-(t)}{4(1-\eta_n)}, 
  \nonumber\\
  %f_n^\pm(t)&=\left(1\pm\sqrt{1-\eta_n}-\frac{\eta_n}{2}\right)e^{-\kappa(1\mp\sqrt{1-\eta_n})t}.
  f_n^\pm(t)&=\frac{1}{2}\left(1\pm\sqrt{1-\eta_n}\right)^2e^{-\gamma(1\mp\sqrt{1-\eta_n})t}.
\end{eqnarray}
In the limit of vanishing rescaled interaction strength $\eta_n$  the purity
tends to one, as it can be noted that $\lim\limits_{\eta_n\to 0}f_n^\pm(t)=1\pm1$. 

In figure \ref{plotD} we have plotted the purity $P(t)$ and atomic excitation probability $p_{\rm e}(t)$ 
for different values of the dephasing parameter $\gamma$. The stationary state of the atomic sub-system 
is the totally mixed state. This explains the drop of purity as a function of time
and the asymptotic value $1/2$ of the excitation probability. 
However, the basic effect is evident, for increasing values of $\gamma$ the purity and excitation 
probability have a slower decay.  
Closer inspection shows that the two quantities are, in this case, closely related as $p_{\rm e}(t)$ has to take 
the value $1/2$ for purity to reach the minimal value of $1/2$.
%Such a relation will not hold for non-solvable problems, yet the general tendencies are the one we expect.

We close this section by pointing out that the expressions 
in Eq. \eref{evalLD} are valid for any value of the parameters as long as $\delta= 0$. However
note that when $\eta_n>1$, the eigenvalues are complex and therefore the
imaginary part gives rise to oscillations in the dynamics, as 
corroborated by the green (dotted)
curves in Fig. 2, where  $\eta_1=16$. This behaviour shows that for small values of $\gamma$ or comparable with $g\sqrt n$, the  
 cavity influences the evolution of the atom through the Hamiltonian interaction in Eq. \eref{Hamiltonian}. 

\section{Photon losses} 
A more common and realistic scenario is the case of photon losses from the cavity. This effect 
can be incorporated by describing the dynamics in terms of the Kossakowski-Lindblad  master equation
\begin{equation}
  \dot\rho=\cL_{\rm l}\rho=
  -i\left[H,\rho\right] +\kappa \cD[a]\rho.
  \label{masterEq}
\end{equation}
Here we have used again  the dissipator defined in Eq. \eref{dissipator}, but in this case we have considered
the Lindblad operator $A=a$ which describes the damping mechanism due to photon losses in the cavity.
The diagonalization of the Liouville operator $\cL_{\rm l}$ in Eq. \eref{masterEq} allows to
evaluate the time evolution of any given initial state. It was noted in Ref. \cite{Torres2014} 
that this procedure can be based on the diagonalization of the non-Hermitian Hamiltonian 
$K=H-i \kappa a^\dagger a/2$ which has the following eigenvalues
\begin{equation}
  \varepsilon_j^{(n)}=
  \frac{2\delta+i\kappa -i2n\kappa}{4}
%  +(-1)^j\sqrt{g^2n+\tfrac{\left(2\delta+i\kappa\right)^2}{16}},
+
\frac{2\delta+i\kappa}{4}
(-1)^j
\sqrt{1+\chi_n}
,
  \label{eivalJC}
\end{equation}
with $j=1,2$ for $n>0$, $j=1$ for $n=0$, and  where we have introduced the rescaled interaction
strength 
\begin{equation}
\chi_n=16g^2n/(2\delta+i\kappa)^2.
  \label{chi}
\end{equation}
In fact, the eigenvalues of $\cL_{\rm l}$ are given by the simple addition of
two eigenvalues of $K$, i.e., 
\begin{equation}
  \lambda_{j,k}^{(n,m)}=
-i(\varepsilon_j^{(n)}-\varepsilon_k^{(m)\ast}),
\end{equation}
%$-i(\varepsilon_j^{(n)}-\varepsilon_k^{(m)\ast})$, 
with $k\in\{1,2\}$ for $m\in \mathbb{N}_+$ and $k=1$ for $m=0$ (as for $j$ and $n$) \cite{Torres2014}. 
Therefore, analyzing the eigenvalues of $K$ in Eq. \eref{eivalJC} implies doing the same for the Liouvillian $\cL_{\rm l}$.
The form of the solution is similar to the eigenvalues of Eq. \eref{evalLD} and an analogous
expansion to the one in Eq. \eref{expevalD} can be performed.  In this way one is able to
analyze the eigenvalues in terms of an expansion in powers of $\chi_n$ 
that appears in the radical of Eq. \eref{eivalJC}. 
This procedure leads to the following expression 
\begin{equation}
  \varepsilon_j^{(n)}=-i\frac{n\kappa}{2}
  +\frac{2\delta+i\kappa}{2}\left( 
  \delta_{j,2}+\frac{(-1)^j}{4}\chi_n+\mathcal{O}[\chi_n^2]\right),
  \label{expeval}
\end{equation}
where $\delta_{i,2}$ is the Kronecker delta, 
$j=1,2$ for $n>0$, and $j=1$ for $n=0$. 
Note that the parameter $\chi_n$  tends to zero with increasing values of $\kappa$ if 
$g$, $n$ and $\delta$ remain finite. 
It follows from the expansion in Eq. \eref{expeval} that the eigenvalues of
the Liouville operator $\cL_{\rm l}$ are insensitive to the coupling strength for sufficiently large values of the photon
decay rate $\kappa$. In this case we also get an indication that the atomic system is ``protected'' from the 
cavity, this time due to photon losses. 
The result holds for arbitrary values of excitation number $n$ and
in this case for all the eigenvalues of the Liouville operator $\cL_{\rm l}$ as they are a sum of two eigenvalues
of Eq. \eref{eivalJC}. In this example involving photon losses, the detuning can also be taken into account thanks
to the simple diagonalization of $\cL_{\rm l}$ that breaks down to the diagonalization of $2\times 2$ matrices.

\begin{figure}[t!]
\begin{center}
\includegraphics[width=.45\textwidth]{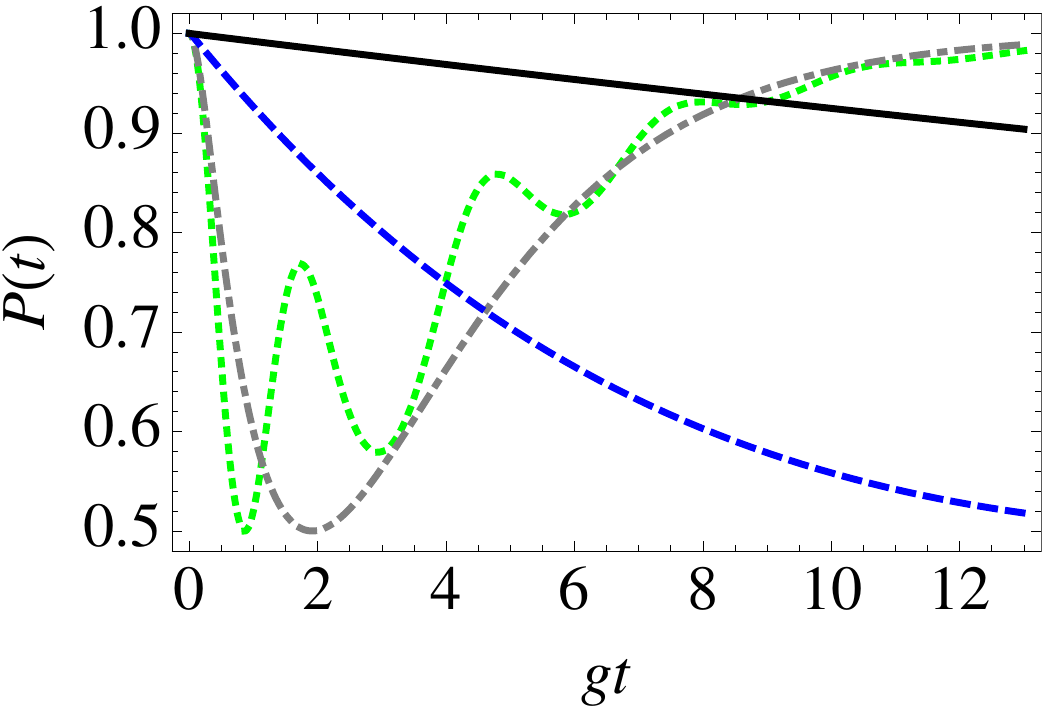}
\includegraphics[width=.45\textwidth]{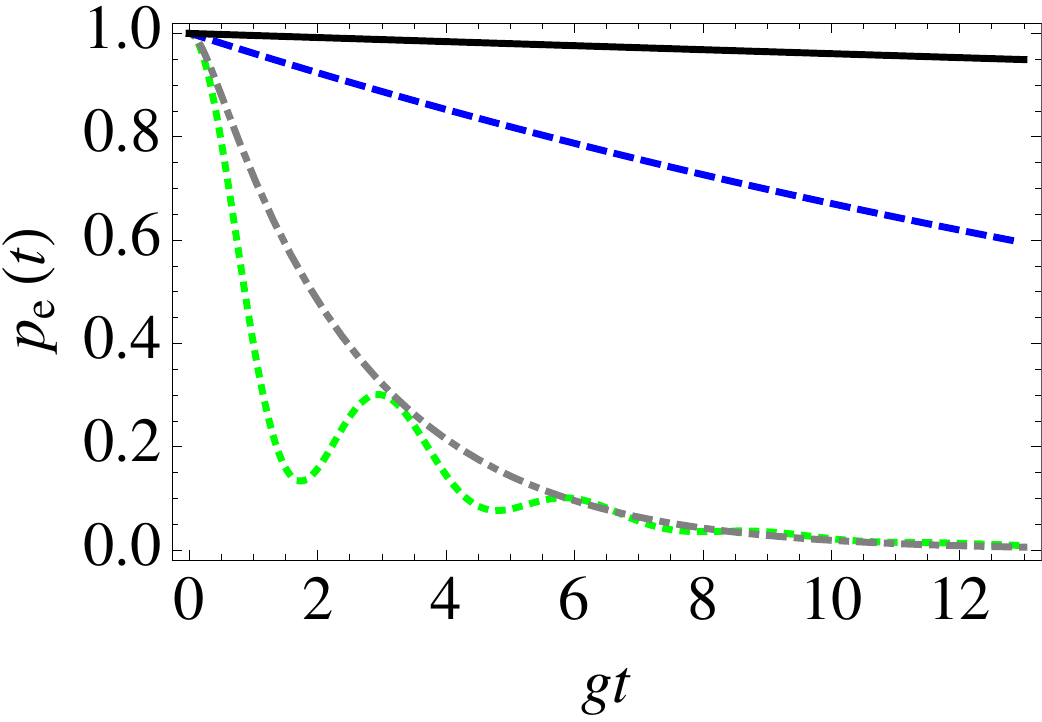}
\caption{\label{plot} Left (right) plot: atomic purity $P(t)$ (excited state population  $p_{\rm e}(t)$) as a function of time in units of $1/g$ for different values of the cavity decay rate $\kappa$. 
The initial state is an excited atom and
an empty cavity subject to the dynamics of Eq. \eref{masterEq}.
Dotted (green) line $\kappa/g=1$, dotted-dashed (grey) line $\kappa/g=10$,
dashed (blue) line $\kappa/g=100$, and full (black) line $\kappa/g=1000$. Detuning is $\delta/g=0.8$.
Note: The purity $P(t)$ is plotted in the range of minimum to maximum purity ($[0.5,1]$).}
\end{center}
\end{figure}

Let us now study  the dynamical features of the atomic sub-system.  
Using the approaches of Refs. \cite{Torres2014,Briegel1993},  it is possible to evaluate the eigensystem of $\cL_{\rm l}$ and in turn
to calculate the time evolution of, in principle,  any given initial condition. For the sake of simplicity, we consider
the case where only one excitation is present in the system. Details of the calculations can be found in \ref{appendix}.
We assume an arbitrary pure state of the atom and
an empty cavity in the Fock state $\ket 0$. The initial condition is then given by 
\begin{equation}
  \ket{\Psi_0}=\left(c_\tg\ket{\tg}+c_\te\ket{\te}\right)\otimes\ket 0.
  %=a\ket{0,1}+b\ket{1,2}0
  \label{initial}
\end{equation}
The corresponding reduced density matrix of the atomic sub-system for this particular initial state is found
to be
\begin{eqnarray}
\fl  \rho_{\rm at}(t)=
  \left(1-|c_\te f(t)|^2\right)\ketbra{\tg}{\tg}
  +|c_\te f(t)|^2
  \ketbra{\te}{\te}
  +
  c_\te c_\tg^\ast f(t) \ketbra{\te}{\tg}  
  +c_\te^\ast c_\tg f^\ast(t)\ketbra{\tg}{\te}
  \label{reducedrho}
\end{eqnarray}
with the complex function
\begin{equation}
  f(t)=
  \frac{4g^2e^{-i\varepsilon_2^{(1)}t}+(i\kappa+
  2\varepsilon^{(1)}_1)^2e^{-i\varepsilon_1^{(1)}t}}
  {4g^2+(i\kappa+2\varepsilon^{(1)}_1)^2}.
  \label{efe}
\end{equation}
It can be seen that $\lim\limits_{\chi_0\to\infty}f(t)=e^{-i\delta t}$, as from
Eq. \eref{expeval} it follows that in this limit the eigenvalues of the non-Hermitian Hamiltonian  $K$
tend to: $\varepsilon_{1}^{(1)}\to -i\kappa/2$ and $\varepsilon_{2}^{(1)}\to \delta$.
Therefore, for large values of the damping parameter $\kappa$ with respect to $g$,
the atom evolves freely under the influence of the free Hamiltonian $\delta\sigma_+\sigma_-$. This can be noted from Eq. \eref{reducedrho} or by evaluating
the population of the atomic excited state 
\begin{equation}
  p_{\rm e}(t)={\rm Tr}\left\{\sigma_+\sigma_-\rho_{\rm at}(t)\right\}=|c_\te f(t)|^2.
\end{equation}
By tracing the square of the atomic density matrix one can 
find that the atomic purity as a function of time is given by the expression 
\begin{equation}
  P(t)=1+2|c_\te|^4|f(t)|^2\left(|f(t)|^2-1\right).
\end{equation}
The form of $P(t)$ makes evident that the atom remains pure for longer times 
as $\kappa$ increases with respect to $g$ and remains completely pure in the limit $\chi_0\to 0$ ($\kappa \gg g$).

Figure \ref{plot} shows the purity $P(t)$ and excited state
population $p_{\rm e}(t)$ of the atom as a function of time for different values of the decay parameter $\kappa$.
The asymptotic behaviour is explained by the knowledge of the steady state, which is the atom in the ground state and an empty cavity. The reason for this is the photon losses that drain all excitations  in the system.
The asymptotic steady state is actually a pure state which explains the
re-emergence of the purity for
large values of the interaction time. The excitation probability drops to zero to never revive also in accordance
with the steady state involving the atom in the ground state.
The important feature is, however, that all of this happens at larger time scales with 
increasing values of the photon losses $\kappa$. This means that very strong coupling of the cavity
to its environment protects the atomic state. In the limit this state is frozen, i.e we again find a quantum Zeno effect.

We close this section by commenting on previous findings related this part of our work.
Similar studies have been considered for a high finesse cavity coupled to a leaky cavity, were 
first numerical results
where given by Imamo\~glu \cite{Imamoglu} followed by analytical investigations of Nemes \cite{Nemes}. 
A two level atom coupled to the continuum of modes was investigated by 
Kofman and Kurizki \cite{Kurizki}, who found analytical results
for the decay of the excited state population including  interruptions of the unitary dynamics
in  line with the quantum Zeno effect. 
In all these three cases \cite{Imamoglu,Nemes,Kurizki}, the authors find systematically that increasing
the leakage of the cavity slows the
decay of the central system. Note this is true with or without interruptions of the unitary evolutions,
i.e., with or without repeated measurements. This implies that the obstruction
of the decay in the atom by increasing leakiness of the cavity is not due to a quantum Zeno effect although 
it is enhanced by it (see Fig. 1 in Ref. \cite{Kurizki}).

\section{Conclusions}
With the dephasing Lindblad operator we have a very simple example, 
where we can see the entire development of decoherence as a function of time 
for different parameters. The scaling behaviour is readily established and we see, 
that  the same equation describes the improvement of decoherence from 
the perturbative all the way to the strong coupling regime, as could be hoped from the 
rather general analytic results implicit in references \cite{Zanardi} and \cite{MGS, GMS}. 
Yet in the perturbative regime for $\kappa$ it seemed that for chaotic environments in the 
Fermi golden rule regime, i.e. with exponential coherence decay, the effect of the far 
environment was no longer noticeable. The latter effect is not seen in our system, where
we always have a preserving effect of the coherence in the central system.

For the case of loss, the fact that we always return to a pure state is trivial, as the
vacuum is the steady state,
but the fact that the initial decoherence slows down as we increase the coupling to 
the far environment is non-trivial.
We thus see, that while the same equation governs the system, the qualitative explanation 
using the quantum Zeno effect will only describe the strong coupling limit, as for weak 
coupling we have complete decoherence and later recoherence while there is a transition to another state. 
This becomes most clearly visible at the opposite end of the coupling range.
Here we see oscillations both in the occupation number and in purity. The unitary Hamiltonian which causes an oscillation of the excitation between the spin and the oscillator is effective. For stronger coupling this dynamics loses importance until we reach a total freeze of dynamics including purity;
%This is manifested in the purity as an oscillatory behaviour that precludes the interpretation as a quantum 
%Zeno effect; 
this is in agreement with the findings of \cite{MGS}. 
There the protection of coherence by decoherence of the environment was shown in a weak coupling 
regime by linear response considerations, which also preclude a quantum Zeno effect.

Thus our two simple models go a long way toward explaining what is going on in the matter 
of decoherence of a near environment protecting the central system. Yet, as is to be expected, 
some aspects of more realistic systems are not covered by the model behaviour.

\ack
J.M.T. acknowledges support by the DFG as part of the CRC 1119 CROSSING and by CONACyT through the 
program Repatriaciones 2016.
T.H.S. whishes to emphasize that he got the basic idea for this work from his late 
friend M.C. Nemes. He also acknowledges financial support from CONACyT research grant 
219993 and PAPIIT-DGAPA-UNAM research grant IG100616.

\appendix
\section{Eigenvalues of $L$}
\label{appendixL}
In this appendix we present the exact eigenvalues of the $4\times4$ matrix $L$ in Eq. \eref{Lmatrix}. The eigenvalues are also 
 roots of the fourth order characteristic polynomial of $L$:
\begin{equation}
  Q(z)=z\left(4g^2n\gamma+(4g^2n+\gamma^2+\delta^2)z+2\gamma z^2+z^3\right).
  %\equiv a_0+a_1 x+a_2x^2+a_3x^3.
  \label{Qpol}
\end{equation}
One of the solutions, namely $Q(0)=0$, can be immediately identified by inspection of  Eq.  \eref{Qpol}. 
The rest of the eigenvalues are roots of the third order polynomial $Q(z)/z$.
Using the solution of the cubic equation in Ref. \cite{AS}, the  four zeros of $Q(z)$ and  eigenvalues of $L$ can be written as
\begin{eqnarray}
  z_0=0,\quad z_1=-\frac{2}{3}\gamma+\left( s-\frac{q}{s}\right),\nonumber\\
  z_\pm=-\frac{2}{3}\gamma-\frac{1}{2}\left( s-\frac{q}{s}\right)
  \pm i\frac{\sqrt3}{2}\left(s+\frac{q}{s}\right),
  \label{zvals}
\end{eqnarray}
where we have introduced the following definitions
\begin{eqnarray}
  q=-\frac{\gamma^2-3\delta^2-12g^2n}{9},\,\quad r=\gamma\frac{\gamma^2+9\delta^2-18g^2n}{27},
  \nonumber\\
  s=\left(r+\sqrt{q^3+r^2}\right)^{1/3}.
  \label{zdefs}
\end{eqnarray}
For the sake of simplicity we restrict our discussion to the case $\delta=0$ in the main text.

\section{Eigensystem of $\cL_{\rm l}$}
\label{appendix}
Here we present  details of the calculations using the eigensystem of the $\cL_{\rm l}$.
The non-Hermitian Hamiltonian $K=H-i\kappa a^\dagger a/2$ can be  
diagonalized in blocks in the basis $\{\ket{n,j}\}$ with the states
\begin{eqnarray}
\ket{n,1}&=\ket{n}\otimes\ket{\tg}, \quad n\ge 0
\nonumber\\
\ket{n,2}&=\ket{n-1}\otimes\ket{\te}, \quad n>0.
\label{basisJC}
\end{eqnarray}
The number state $\ket{n}$ describes a situation of $n$ photons
in the cavity, while  $\ket{\te}$ and $\ket{\tg}$ stands for the excited and ground state  of the atom.
The explicit form of the blocks of $K$ is given by 
\begin{eqnarray}
  &K^{(n)}=
  \left(
  \begin{array}{cc}
    -i\frac{ n \kappa}{2}
    & g\sqrt{n}
    \\
    g\sqrt{n}&
    \frac{2\delta-i(n-1)\kappa}{2}
  \end{array}
  \right),
\label{blocksJC}
\end{eqnarray}
for $n>0$ and $K^{(0)}=0$. The eigenvalues are given in Eq. \eref{eivalJC}.
The diagonalization of the matrices $K^{(n)}$  can be accomplished with the transformation 
$R^{(n)\top} H^{(n)}R^{(n)}$, with
\begin{eqnarray}
R^{(n)}&=
  \left(
  \begin{array}{cc}
    \cos\theta_n&-\sin\theta_n\\
    \sin\theta_n&\cos\theta_n
  \end{array}
  \right),
  \label{RJC}
\end{eqnarray}
with 
$\theta_n=\arctan\left[(2\varepsilon_1^{(n)}+i n\kappa)/2 g\sqrt{n}\right]$.
The right and left eigenvectors of $K$ are given by
\begin{eqnarray}
  \ket{r_j^{(n)}}=
  \sum_{j=1}^2R_{k,j}^{(n)}\ket{n,k},\quad
  \bra{q_j^{(n)}}=
  \sum_{j=1}^2R_{k,j}^{(n)}\bra{n,k}
  \label{eivec}
\end{eqnarray}
for $n>0$ and the singlet $\ket{0,1}$ for $n=0$.
It has been shown in Ref. \cite{Torres2014} that the full eigensystem of the Liouville operator $\cL_{\rm l}$
in Eq. \eref{masterEq} can be constructed from the eigensystem of the non-Hermitian Hamiltonian $K$.
With the knowledge of the full set of right (left) eigenvectors $\hat\rho_\lambda$ 
($\check\rho_\lambda$) (labeled with the corresponding eigenvalue $\lambda$), one is able to evaluate the time evolution of any
given initial condition $\rho_0$ as 
\begin{eqnarray}
  \rho(t)=e^{\cL_{\rm l} t}\rho_0=\sum_\lambda {\rm Tr}\left\{\check\rho_\lambda^\dagger\rho_0\right\}
  e^{\lambda t} \hat\rho_\lambda.
  \label{tevol}
\end{eqnarray}
For initial states  $\ketbra{\Psi_0}{\Psi_0}$ 
of Eq. \eref{initial}, the only contribution to \eref{tevol}
is given by the following set of $7$ right  
eigenvectors of $\cL_{\rm l}$
\begin{eqnarray}
\fl\hat\rho_{1,1}^{(0,0)}=\ketbra{0,1}{0,1},
\quad
\hat\rho_{j,1}^{(1,0)}=\ketbra{r^1_j}{0,1},\quad
\hat\rho_{j,k}^{(1,1)}=\ketbra{r_j^1}{r_k^1}-\braket{r^1_k}{r^1_j}\ketbra{0,1}{0,1}.
\end{eqnarray}
The corresponding left eigenvectors are $\check\rho_{1,1}^{(0,0)}=\mathbb{I}$,
$\check\rho_{j,1}^{(1,0)}=\ketbra{q^1_j}{0,1}+\dots$ and
$\check\rho_{j,k}^{(0,1)}=\ketbra{q_j^1}{q_k^1}+\dots$, where the dots indicate a series of terms that we omit 
as they do not contribute to initial states describing one excitation in the system.
The corresponding eigenvalues are
$\lambda^{(0,0)}_{1,1}=0$, 
$\lambda^{(1,0)}_{j,1}=-i\varepsilon^{(1)}_j$ and
$\lambda^{(1,1)}_{j,k}=-i\left(\varepsilon^{(1)}_j-\varepsilon^{\ast(1)}_k\right)$.
With this subset of the eigensystem, it is possible to write the time evolution of the initial state in Eq. 
\eref{initial} as 
\begin{eqnarray}
  \fl\rho(t)=\ketbra{0,1}{0,1}+\sum_{j,k=1}^2\bra{q^1_j}\rho_0\ket{q_k^1}
  e^{\lambda^{(1,1)}_{j,k}t}
  \hat\rho_{j,k}^{(1,1)}
  +\sum_{j=1}^2\left(
  \bra{q_j^1}\rho_0\ket{0,1}e^{\lambda_{j,1}^{(1,0)}t}\hat\rho_{j,1}^{(1,0)}
  +{\rm H.c.}
  \right),
\end{eqnarray}
with $\rho_0=\ketbra{\Psi_0}{\Psi_0}$. From Eqs. \eref{initial} and \eref{eivec} it follows that
$\braket{q_j^1}{\Psi_0}=c_\te R_{2,j}^{(1)}$ and $ \braket{\Psi_0}{0,1}=c_\tg^\ast$.
By taking this into account and tracing over the photonic degrees of freedom on finds
the reduced density matrix of the atomic system  
\newpage
\begin{eqnarray}
  \rho_{\rm at}(t)
  =&
  \ketbra{\tg}{\tg}
  +\sum_{j=1}^2\left(
  c_\te c_\tg^\ast (R_{2,j}^{(1)})^2e^{\lambda_{j,1}^{(1,0)}t}\ketbra{\te}{\tg}  
  +{\rm H.c.}
  \right)
  \nonumber\\
  &
%  +b^\ast a (R_{2,j}^{(1)\ast})^2e^{\lambda_{1,j}^{(-1,0)}t}\ketbra{g}{e}
  +|c_\te|^2
  \sum_{j,k=1}^2 
  \left(R_{2,j}^{(1)}R_{2,k}^{(1)\ast}\right)^2
  e^{\lambda_{j,k}^{(1,1)}t}
  \left(
  \ketbra{\te}{\te}
  -\ketbra{\tg}{\tg}
  \right).
\end{eqnarray}
Finally, the expression in Eq. \eref{reducedrho} is obtained by identifying that
\begin{equation}
  f(t)=
 \sum_{j=1}^2
  R_{2,j}^{(1)}e^{-i\varepsilon_j^{(1)}t}.
  \label{efe0}
\end{equation}

\end{document}